\documentclass[12pt]{article}
\usepackage{hyperref}
\usepackage{amsmath}
\usepackage{graphicx}
\usepackage{amssymb} 
\usepackage{color}
\usepackage{soul} 
\definecolor{mypurple}{rgb}{0.71,0.02,1}

\def\be{\begin{equation}}
\def\ee{\end{equation}}
\def\bea{\begin{eqnarray}}
\def\eea{\end{eqnarray}}
\def\bi{\begin{itemize}}
\def\ei{\end{itemize}}

\date{}

\title{Equivalent forms of Dirac equations in curved spacetimes and generalized de Broglie relations}

\author{
Mayeul Arminjon\,$^{1}$ and Frank Reifler\,$^2$\\
$^1$ \small\it Laboratory ``Soils, Solids, Structures, Risks'' \\
\small\it (CNRS and Universit\'es de Grenoble: UJF, G-INP),\\
\small\it BP 53, F-38041 Grenoble cedex 9, France.\\
\small\it $^2$ Lockheed Martin Corporation, MS2 137-205,\\ 
\small\it 199 Borton Landing Road, Moorestown, New Jersey 08057, USA.
} 

\begin{document}
\maketitle

\begin{abstract}
\noindent  One may ask whether the relations between energy and frequency and between momentum and wave vector, introduced for matter waves by de Broglie, are rigorously valid in the presence of gravity.  In this paper, we show this to be true for Dirac equations in a background of gravitational and electromagnetic fields. We first transform any Dirac equation into an equivalent canonical form, sometimes used in particular cases to solve Dirac equations in a curved spacetime. This canonical form is needed to apply the Whitham Lagrangian method. The latter method, unlike the WKB
method, places no restriction on the magnitude of Planck's constant to obtain wave packets, and furthermore preserves the symmetries of the Dirac Lagrangian. We show using canonical Dirac fields in a curved spacetime, that the probability current has a Gordon decomposition into a convection current and a spin current, and that the spin current vanishes in the Whitham approximation, which explains the negligible effect of spin on wave packet solutions, independent of the size of Planck's constant. We further discuss the classical-quantum correspondence in a curved spacetime based on both Lagrangian and Hamiltonian formulations of the Whitham equations. We show that the generalized de Broglie relations in a curved spacetime are a direct consequence of Whitham's Lagrangian method, and not just a physical hypothesis as introduced by Einstein and de Broglie, and by many quantum mechanics textbooks.

\end{abstract}

\noindent                              03.75. - b   Matter waves

\noindent                              04.62. + v~  Quantum fields in curved spacetime

\section{Introduction}

Some might argue that quantum mechanics became a universal theory of matter when, at the last Einstein-Bohr debate at the Solvay Conference in 1930, Einstein proposed a weight measurement to observe unobtrusively a particle decay in order to contradict the Heisenberg energy-time uncertainty relation \cite{1.}.  After nearly being defeated by Einstein in the debate, Bohr surprisingly countered with a general relativistic gravitational argument.  Henceforth, the relation between gravity and quantum mechanics was to become an important question in fundamental physics.

It could also have been questioned whether the relations between energy and frequency and between momentum and wave vector, introduced for matter waves six years earlier by de Broglie \cite{1.}, were rigorously valid in a general curved spacetime.  This question can be shown to be equivalent to the question of whether sufficiently small wave packets travel along classical paths consistent with the de Broglie relations.  Recall that a wave packet is a wave whose amplitude, frequency, and wave vector vary slowly over a region of spacetime comparable to a period or wave length.  (E.g., for an electron traveling at half the speed of light, the wave length is approximately 5 $\times$ 10${}^{-}$${}^{12}$ m, and a typical wave packet has dimensions 10${}^{-}$${}^{6}$ m \cite{2.}.)  The de Broglie relations can be observed for such wave packets. 

The well known WKB approximation is commonly used to derive wave packet approximations in quantum mechanics (e.g., \cite{Audretsch1981A}).  The WKB approximation is based on taking the limit as a physical constant, namely Planck's constant $\hbar $, approaches  zero. 
\footnote{\
More generally, a common conception is that classical physics emerges from quantum mechanics in the limit as Planck's constant $\hbar $ approaches zero. However, the limit $\hbar \rightarrow 0$ ``is not well defined mathematically unless one specifies what quantities are to be held  constant during the limiting process" \cite{Ballentine2001}. It is interesting to note that the most classical behaving Gaussian wave functions, the coherent states of the ordinary harmonic oscillator, whose expected position and momentum obey classical equations by Ehrenfest's theorem, do not resemble wave packets in the limit $\hbar \rightarrow 0$ \cite{Park1990}. 
}
However, the spin connection in the Dirac equation of a curved spacetime has no effect in the first WKB approximation (i.e., the one retaining only the zero order term in $\hbar $) \cite{Audretsch1981A}. The assumption that the spin connection can be neglected, as it would be in a first WKB approximation, is unnecessary
and is too strong for many applications in a curved spacetime, or even in a Minkowski spacetime with arbitrary coordinates. Note that throughout this paper, except for the brief description of a post Newtonian approximation in Section \ref{WhithamMethod}, we may set both the speed of light $c$ and Planck's constant $\hbar $  equal to one.

The Whitham approximation \cite{12.}, which we adopt in this paper, places no restriction on Planck's constant. To implement the Whitham approximation and to show that it leads to propagation along classical paths, we will first show in Section \ref{Canonical Form} that any Dirac equation in a curved spacetime can be transformed into an equivalent canonical form known in the literature as the ``local representation" \cite{3.}, \cite{4.}.  In general, transformation to equivalent canonical form is a necessary step to simplify a Dirac equation so that propagation along classical paths can be derived.  It will be evident in Theorem 1 of Section \ref{Canonical Form} that Planck's constant $\hbar $ does not appear in the transformations mapping Dirac equations to their equivalent canonical forms \cite{3.}, \cite{4.}.  Previously, these canonical forms, or ``local representations" as they are called in the literature, have only been discussed in the special case of orthogonal coordinates \cite{3.}, \cite{4.}, \cite{5.}.

Then in Section \ref{WhithamMethod}, with each Dirac equation transformed into equivalent canonical form, we apply Whitham's Lagrangian method \cite{12.} to derive wave packets in general curved spacetimes.  Whitham's method preserves the symmetries of the Lagrangian, and in particular, the Whitham wave packet equations conserve the probability current.  We also show that generalized de Broglie relations, as well as COW and Sagnac type terms \cite{10.}, emerge from the Whitham equations.  It will become clear in Sections 2 and 3 that for every Dirac equation transformed into equivalent canonical form, the generalized de Broglie relations have no other meaning than the fact that sufficiently small wave packets propagate along classical paths in a background of gravitational and electromagnetic fields.  This is what is observed in experiments 
\footnote{  Note, however, that the electron's magnetic moment, predicted by the Dirac equation, is not contained in the wave packet approximation.  Indeed, quoting from Ref. \cite{13.}:  ``The uncertainty principle, together with the Lorentz force, prevents spin-up and spin-down electrons from being separated by a macroscopic field of the Stern-Gerlach type.''  In practice, wave packet splitting in Stern-Gerlach experiments is only observed using neutral atoms or molecules, which are undisturbed by the Lorentz force \cite{Ballentine2001}.  In Section \ref{WhithamMethod}, the wave packet approximation is expressed by neglecting in the Lagrangian the variation in the amplitude of the wave function as compared to the variation of its phase.  This leads to wave packet equations which do not involve the electron's magnetic moment.    
} 
and therefore more physically precise than the statement often made that particles with a given energy and momentum possess a frequency and wave vector given by the de Broglie relations.  In fact, the generalized de Broglie relations are a \textit{direct consequence} of Whitham's method applied to each Dirac equation transformed into equivalent canonical form, and not just a physical \textit{hypothesis} as introduced by Einstein and de Broglie, and by many quantum mechanics textbooks.

In the WKB approximation of the standard Dirac equation in a curved spacetime, classical trajectories are derived from the Gordon decomposition of the probability current $J^\mu = J^\mu _c + J^\mu _s$ into a convection current $J^\mu _c$ and a spin current $J^\mu _s$ \cite{Audretsch1981A}. In Section \ref{WhithamMethod}, we also prove the existence of the Gordon decomposition for Dirac equations transformed into canonical form. We further show that in the Whitham approximation, the spin current $J^\mu _s$ vanishes, which explains the negligible effect of spin on wave packet solutions, independent of the size of Planck's constant $\hbar $.

It is also clear that the canonical forms (or ``local representations'') of the Dirac equations, while not unique, are the preferred representations to understand certain phenomena associated with the Dirac equations in a curved spacetime, particularly, the emergence of classical physics and its conservation laws in a quantum world.
Section \ref{Classical-Quantum} concludes this paper with a discussion of the classical-quantum correspondence in a curved spacetime based on both Lagrangian and Hamiltonian formulations of the Whitham equations.  In this section we also include results from a previous analysis of the classical-quantum correspondence \cite{15.}, which can be applied to the canonical forms of Dirac equations in a curved spacetime considered in this paper.

\section{New Representations of the Dirac Equation in a Curved Spacetime and their Equivalent Canonical Forms}\label{Canonical Form}

Shortly after Dirac discovered his celebrated four component wave equation:

\be 
\gamma ^{\mu } \partial _{\mu } \Psi = -\frac{imc}{\hbar }  \Psi, 
\label{GrindEQ__1_}
\ee
together with its conserved probability current:
\be J^{\mu } = c \Psi ^{+} A\gamma ^{\mu }  \Psi, 
 \label{GrindEQ__2_} 
\ee
his equation was studied in its widest representations for a Minkowski spacetime \cite{6.}, \cite{16.}.  

In Eq. \eqref{GrindEQ__1_}, the Dirac field $ \Psi  $ is a four component complex function of spacetime coordinates $ x^{\mu }  $, $ \mu   =  0 , 1 , 2, 3 $, whose partial derivatives with respect to $ x^{\mu }  $ are denoted as $ \partial _{\mu } \Psi  $.  The Dirac gamma matrices $ \gamma ^{\mu }  $, acting on $ \Psi  $, satisfy the anticommutation formula:
\be
\gamma ^{\mu } \gamma ^{\nu }   +   \gamma ^{\nu } \gamma ^{\mu } = 2 \eta ^{\mu \nu }  {\bf 1}_{{\bf 4}}, \label{GrindEQ__3_}
\ee
where $ \eta ^{\mu \nu }  $ is the inverse of the Minkowski metric tensor $ \eta _{\mu \nu }  $, and $ {\bf 1}_{{\bf 4}}  $ denotes the identity matrix acting on the Dirac field $ \Psi  $.  The mass, the speed of light, and Planck's constant are denoted by $ m $, $ c $, and $ \hbar  $, respectively.  Repeated indices are summed.

In Eq. \eqref{GrindEQ__2_}, $ \Psi ^{+}  $ denotes the complex conjugate transpose (Hermitian conjugate) of the Dirac field $ \Psi  $, and $ A $ is a hermitizing matrix for the Dirac gamma matrices $ \gamma ^{\mu }  $.  That is, \cite{6.}, \cite{16.}:

\noindent  
\be\begin{array}{l} {A^{+} = A,} \\ {} \\ {\gamma ^{\mu +} = A \gamma ^{\mu } A^{-1} }, \end{array} \label{GrindEQ__4_}\ee

\noindent where $ \gamma ^{\mu +}  $ and  $ A^{+}  $ denote the Hermitian conjugates of the matrices $ \gamma ^{\mu }  $ and  $ A $, respectively.  The hermitizing matrix $ A $ is uniquely determined by the matrices $ \gamma ^{\mu }  $ up to a nonzero real scalar multiple \cite{6.}. 

For a Minkowski spacetime, assuming that $ \left( \gamma ^{\mu } , A \right) $ are chosen to be constant matrices satisfying Eqs. \eqref{GrindEQ__3_} and \eqref{GrindEQ__4_}, every solution of Eq. \eqref{GrindEQ__1_} satisfies the Klein-Gordon equation, and the probability current $ J^{\mu }  $ in Eq. \eqref{GrindEQ__2_} is then also conserved.  That is, Eqs. \eqref{GrindEQ__3_} and \eqref{GrindEQ__4_} are the only conditions that the constant matrices $ \left( \gamma ^{\mu } , A \right) $ need satisfy.      

The ``coefficient matrices'' $ \left( \gamma ^{\mu } , A \right) $ in Eqs. \eqref{GrindEQ__1_} and \eqref{GrindEQ__2_} are far from unique.  Given a Dirac field $ \Psi  $, and any set of constant coefficient matrices $ \left( \gamma ^{\mu } , A \right) $ satisfying Eqs. \eqref{GrindEQ__3_} and \eqref{GrindEQ__4_},  they may be transformed by any constant complex $ 4\times 4 $ matrix $ S $ as follows:

\noindent 
\be\begin{array}{l} {\widetilde{\Psi }\quad  =\quad S^{-1}  \Psi, } \\ {} \\ {\widetilde{\gamma }^{\mu } = S^{-1} \gamma ^{\mu }  S,} \\ {} \\ {\widetilde{A}\quad  =\quad S^{+} A S.} \end{array}\label{GrindEQ__5_}
\ee

\noindent Such a transformation $ S $ is called a ``similarity transformation'' or a ``spin-base transformation'', the latter referring to simply a change of basis for the four components of the Dirac field $ \Psi  $.  It is straightforward to see that Eqs. \eqref{GrindEQ__1_} $-$ \eqref{GrindEQ__4_} are invariant under all similarity transformations $ S $ by Eq. \eqref{GrindEQ__5_}.  Thus, in the widest sense no restriction to a smaller group of transformations was deemed necessary in the early development of the Dirac equation \cite{6.}, \cite{16.}. 

\noindent 

\noindent Nevertheless, further choices were necessary when the Poincar\'e group of coordinate transformations of the Dirac equation was considered.  At least three possibilities have been considered for Poincar\'e coordinate transformations in a Minkowski spacetime as follows:  

\noindent 

\bi
\item  A) The Dirac field $ \Psi  $ transforms as a quadruplet of complex scalars under a coordinate transformation.  \cite{17.}, \cite{18.}, \cite{19.}, \cite{20.}\\ 

\item  B) The Dirac field $ \Psi  $ transforms as a complex four-vector $ \Psi ^{\mu }  $ under a coordinate transformation.  \cite{21.}\\ 

\item  C) The Dirac field $ \Psi  $ transforms as a quadruplet of complex scalars under a coordinate transformation, which is then followed by a similarity transformation.  (The combined transformation leaves the constant coefficient matrices $ \left(\gamma ^{\mu } , A\right) $ invariant.)  \cite{6.}\\
\ei

\noindent For a Minkowski spacetime with Poincar\'e coordinate transformations, all three possibilities may be considered.  However, with general coordinate transformations, as required for a curved spacetime, only the first two possibilities (A) and (B) exist.  Since in a general spacetime, the possibility (C) does not exist, it was replaced by the possibility (A) in what has become the standard Dirac equation, which was proposed independently by Weyl \cite{17.} and by Fock \cite{18.}, hereafter called the ``Dirac-Fock-Weyl'' (DFW) equation. See Refs. \cite{19.} and \cite{20.}.  Possibility (B) was investigated recently \cite{21.}, which became the impetus for a more general study of Dirac equations \cite{7.}, \cite{8.}, \cite{22.}.

General Dirac fields of type (A) will be said to belong to the Quadruplet Representation of the Dirac theory (or QRD theory).  General Dirac fields of type (B) will be said to belong to the Tensor Representation of the Dirac theory (or TRD theory).  It was recently shown that in an open neighborhood of each spacetime point, every TRD equation is in fact equivalent to a QRD equation and vice versa \cite{7.}.  Since TRD equations are locally equivalent to QRD equations, we will only consider QRD equations in this paper.  Note that there are QRD equations in a curved spacetime that are not locally equivalent to any DFW equation \cite{7.}.     

In a further evolution of the Dirac equation, which applies also to a Minkowski spacetime, the partial derivatives in the Dirac equation \eqref{GrindEQ__1_} were replaced by covariant derivatives $ D_{\mu } = \partial _{\mu }   +  \Gamma _{\mu }  $ where $ \Gamma _{\mu }  $ are four $ 4\times 4 $ complex matrices 
\footnote{\ 
In the case of a Majorana representation of the Dirac field, the coefficient matrices $   \left( \gamma ^{\mu } , A \right) $  are pure imaginary, and the Dirac equation is real.  In this case the spin connection matrices $ \Gamma _{\mu }  $ are real.   
}
, called ``spin connection matrices'', acting on the four components of the Dirac field $ \Psi  $.  At the same time, the Dirac equation was generalized by substituting a general metric $ g_{\mu \nu }  $ of Lorentz signature and determinant $g$ for the Minkowski metric $ \eta _{\mu \nu }  $ in the anticommutation formula \eqref{GrindEQ__3_} for the Dirac gamma matrices:   

\noindent 
\be\gamma ^{\mu } \gamma ^{\nu }   +   \gamma ^{\nu } \gamma ^{\mu } = 2 g^{\mu \nu }  {\bf 1}_{{\bf 4}}. \label{GrindEQ__7_}
\ee
For the results in this paper, mild restrictions must be placed on the metric components $ g_{\mu \nu }  $, namely that $ g_{00}   >  0 $ and the components $ g_{jk}  $ for $ j, k  =  1, 2, 3 $ form a negative definite $ 3\times 3 $ matrix.  Even though these mild conditions hold for almost all spacetime metrics $ g_{\mu \nu }  $ of interest, the G\"odel metric is a notable exception \cite{23.}, \cite{24.}.  

For this generalization, the coefficient matrices $ \left( \gamma ^{\mu } , A \right) $ defining the Dirac equation are augmented to become ``coefficient fields'' $ \left( \gamma ^{\mu } , A , \Gamma _{\mu }  \right) $, which now may vary with the spacetime point.  Then, in order that the Dirac equation \eqref{GrindEQ__1_} be invariant, the transformation equations \eqref{GrindEQ__5_} are augmented with the transformation of the spin connection matrices $ \Gamma _{\mu }  $ under ``local similarity transformations'' $ S $ (i.e., similarity transformations $ S $ that may also vary with the spacetime point 
\footnote{
Because $ S$ depends on the spacetime point, Schl\"utert, Wietschorke, and Greiner call $ S $ a ``local transformation'' \cite{3.}, \cite{4.}.   }
) as follows: \cite{20.}  

\noindent 
\be\begin{array}{l} { \widetilde{\Gamma }_{\mu } = S^{-1} \Gamma _{\mu } S  +  S^{-1} \partial _{\mu } S} \\ {} \\ {\qquad =S^{-1} \left( \partial _{\mu } +\Gamma _{\mu }  \right) S.} \end{array}
\label{GrindEQ__9_}
\ee
Indeed, for this type of transformation, we have the covariant derivatives transforming as: $\widetilde{D}_{\mu }= S^{-1} \circ D_{\mu } \circ S$.  Transformations of the kind given by Eqs. \eqref{GrindEQ__5_} and \eqref{GrindEQ__9_} we will call local similarity transformations ``of the first kind''.  

Local similarity transformations ``of the second kind'' are defined by setting: 
\be
\widetilde{\Gamma }_{\mu } = \Gamma _{\mu }. \label{GrindEQ__11_}
\ee 
For this second kind of transformation, given by Eqs. \eqref{GrindEQ__5_} and \eqref{GrindEQ__11_}, we have the covariant derivatives transforming as: $ \widetilde{D}_{\mu } = D_{\mu } $.  

Two Dirac equations will be said to be ``equivalent'' or ``classically equivalent'' if there exists a local similarity transformation $ \Psi   \to   S^{-1} \Psi  $ of any kind that transforms the solutions of one Dirac equation into the solutions of the other. 
\footnote{ 
The notion of equivalence here is somewhat different than the notion used in Ref. \cite{25.}, where equivalence was applied only to classify the coefficient fields  $ \left( \gamma ^{\mu } , A \right) $ without requiring the existence of a map $ \Psi   \to   S^{-1} \Psi  $ between the solutions of two Dirac equations.    
}   
From Eqs. \eqref{GrindEQ__2_} and \eqref{GrindEQ__5_}, the conserved probability currents for two equivalent Dirac equations are equal.  Hence in any spacetime, scattering experiments will give the same results regardless of the representation of a given Dirac equation.  However, in a first quantized theory, which is our concern in this paper, a local similarity transformation $ S $ may not intertwine with the quantum mechanical operators.  In that case, the operators corresponding to a given observable generally will not have the same eigenvalues before and after the transformation $S$.  This has been shown for the Hamiltonian (or energy) operator in previous work \cite{25.}.  Thus, two Dirac equations that are equivalent as partial differential equations via a local similarity transformation $S$, need not be equivalent with respect to all quantum mechanical operators \cite{25.}, \cite{26.}, \cite{27.}. \\               

The standard ``Dirac Lagrangian'' applies to the DFW equation \cite{19.} and has to be extended to include the coefficient field $A$ in the exact place of the constant hermitizing matrix valid for DFW \cite{25.}:
\be\begin{array}{l} {L= L\left(\Psi, \partial _{\mu } \Psi , x^{\mu }  \right)} \\ {} \\ {= \sqrt{ -g}\  \frac{i\hbar c }{2}\,  {\rm [} \Psi ^{+} A\gamma ^{\mu } \left( D_{\mu } \Psi  \right) -\left( D_{\mu } \Psi \right)^{+} A\gamma ^{\mu } \Psi  + \frac{ 2 mc }{\hbar }\, i\, \Psi ^{+} A\Psi  {\rm ]}.} \end{array}\label{GrindEQ__8_}
\ee
Note that the Lagrangian (\ref{GrindEQ__8_}) is the local expression of a global Dirac Lagrangian based on a general Dirac operator ${\not}\mathcal{D}$ acting on the cross-sections of a spinor bundle ${\sf E}$ over the spacetime. 
\footnote{\ 
A globally defined generalized Dirac Lagrangian has the form:
\be\label{Lagrangian-intrinsic}
L=\ \frac{i}{2}
\left [(\psi,{\not}\mathcal{D}\psi)-({\not}\mathcal{D}\psi,\psi)+2im(\psi ,\psi) \right].
\ee
where ${\not}\mathcal{D}$ is a Dirac operator acting on the cross-sections $\psi $ of a spinor bundle ${\sf E}$ over the spacetime, and $(\ ,\ )$ denotes a hermitizing metric on the fibers of ${\sf E}$.  See Ref. \cite{7.}, Sect. 2.1 and the references therein for the precise definitions.  Once any coordinate chart of the spacetime and any frame field on the spinor bundle have been chosen, one gets the local expression of the global generalized Lagrangian as Eq. (\ref{GrindEQ__8_}). In particular, in Eq. (\ref{GrindEQ__8_}) and in the rest of this paper, $\Psi $ is the column vector made with the components of $\psi $ in the chosen frame field on ${\sf E}$. See Ref. \cite{7.}, Sect. 2.2.  
}
Thus, the Lagrangian (\ref{GrindEQ__8_}) gives a generalized formulation of the Dirac theory for a general Dirac operator ${\not}\mathcal{D}$ on a curved spacetime. We will see in Theorem 1 that the equivalent canonical forms of DFW equations require such a generalization.

The Euler-Lagrange equations for the Lagrangian \eqref{GrindEQ__8_} give the following generalized Dirac equation \cite{7.}, \cite{8.}:  

\noindent 
\be\gamma ^{\mu } D_{\mu } \Psi    +   \frac{1}{2}  A^{-1} D_{\mu } \left(\, A\gamma ^{\mu } \right) \Psi = -\frac{imc}{\hbar }  \Psi,
\label{GrindEQ__14_}\ee

\noindent where 
\be\begin{array}{l} {D_{\mu } \Psi \quad   \equiv \quad \partial _{\mu } \Psi    +   \Gamma _{\mu }  \Psi } \\ {} \\ {D_{\mu } \gamma ^{\nu } \quad  \equiv  \quad \nabla _{\mu } \gamma ^{\nu }   +   \Gamma _{\mu } \gamma ^{\nu }   -  \gamma ^{\nu } \Gamma _{\mu } } \\ {} \\ {D_{\mu } A\quad   \equiv  \quad \partial _{\mu } A   -   \Gamma _{\mu }^{+}  A   -   A \Gamma _{\mu }, } \end{array} 
\label{GrindEQ__15_}\ee

\noindent and we define the Levi-Civita covariant derivatives \textbf{$ \nabla _{\mu }  $ }acting on the Dirac field $ \Psi  $ and the coefficient fields $ \left( \gamma ^{\mu } , A \right) $ as follows:

\noindent 
\be\begin{array}{l} {\nabla _{\mu } \Psi \quad   \equiv \quad \partial _{\mu } \Psi } \\ {} \\ {\nabla _{\mu } \gamma ^{\nu } \quad  \equiv  \quad \partial _{\mu } \gamma ^{\nu }    +   \left\{^\nu _{\rho\, \mu } \right\} \gamma ^{\rho } } \\ {} \\ {\nabla _{\mu } A\quad   \equiv  \quad \partial _{\mu } A} \end{array} 
\label{GrindEQ__16_}\ee

\noindent where $\left\{^\nu _{\rho\, \mu } \right\}$ are the Christoffel symbols belonging to the Levi-Civita connection.  The covariant derivatives $ D_{\mu } $ extend to  $\Psi ^{+}$ by the formula $D_{\mu } \Psi ^{+}= \left( D_{\mu } \Psi \right)^{+} $, and  similarly $\nabla _{\mu } \Psi ^{+}= \left(\nabla _{\mu } \Psi \right)^{+} $.  As usual, covariant derivatives extend to products of fields via Leibniz's rule for differentiating products.      

If  $D_{\mu } \left(A\gamma ^{\mu } \right)= 0$, then the generalized Dirac equation \eqref{GrindEQ__14_} reduces to normal form:

\noindent 
\be
\gamma ^{\mu } D_{\mu } \Psi = -\frac{imc}{\hbar }  \Psi.      
\label{GrindEQ__17_}
\ee

\noindent Normal Dirac equations generalize the DFW property that the coefficient fields $ \left( \gamma ^{\mu } , A \right) $ be covariantly constant:  $ D_{\mu } \gamma ^{\nu }   =  0 $ and $ D_{\mu } A  =  0 $.  One can show that the weaker normality condition $ D_{\mu } \left( A\gamma ^{\mu } \right)  =  0 $ is preserved under all local similarity transformations of the first kind \cite{7.}.  The normality condition is also preserved under all coordinate transformations. 

Thus we have several invariant classes of Dirac equations.  First, we have the class of Dirac equations for which the coefficient fields $ \left( \gamma ^{\mu } , A \right) $ are covariantly constant $-$ that is, $ D_{\mu } \gamma ^{\nu }   =  0 $ and $ D_{\mu } A  =  0 $.  This class contains the DFW equations as a proper subset.  Second, we have the class of Dirac equations for which $ D_{\mu } \left( A\gamma ^{\mu } \right)  =  0 $.  This is the class of normal Dirac equations \eqref{GrindEQ__17_} which contains the first class as a proper subset.  Finally, we have the class of generalized Dirac equations \eqref{GrindEQ__14_} which contains the other two classes.  Each of these classes is invariant under all coordinate transformations and also under all local similarity transformations of the first kind \cite{7.}.  

A non-invariant class of Dirac equations that we will discuss in this paper is the class of the QRD--0 equations, in which the contracted spin connection matrix $ \Gamma \equiv \gamma ^{\mu } \Gamma _{\mu } = 0 $.  See Ref. \cite{7.}, Sect. 3.2.1. Equations in the QRD--0  class (with $ \Gamma = 0$) are generalized Dirac equations which may be written as follows:    
\be
\gamma ^{\mu } \partial _{\mu } \Psi    +   \frac{1}{2}  A^{-1} \nabla _{\mu } \left(\, A\gamma ^{\mu } \right)\Psi =-\frac{imc}{\hbar } \Psi,   
\label{GrindEQ__18_}
\ee
with the Levi-Civita covariant derivatives $ \nabla _{\mu }  $ acting on the coefficient fields $ \left( \gamma ^{\mu } , A \right) $ as previously defined in Eq. \eqref{GrindEQ__16_}.  If $ \nabla _{\mu } \left( A\gamma ^{\mu } \right)  =  0 $, then the QRD--0 equation \eqref{GrindEQ__18_} reduces to the simpler normal form:

\noindent 
\be\gamma ^{\mu } \partial _{\mu } \Psi = -\frac{imc}{\hbar }  \Psi .   
\label{GrindEQ__19_}\ee

\noindent Note that a QRD--0 equation which is both normal as in Eq. \eqref{GrindEQ__19_} and equivalent to a DFW equation is called a ``local representation'' of the DFW equation by other authors \cite{3.}, \cite{4.}, \cite{5.}.   Finding a ``local representation'' for a DFW equation often simplifies the process of deriving its solutions.  Previously, this ``local representation'' of the DFW equation in a curved spacetime in the form of Eq. \eqref{GrindEQ__19_} has only been discussed in the special case of orthogonal coordinates \cite{3.}, \cite{4.}, \cite{5.}.

Given any normal Dirac equation \eqref{GrindEQ__17_}, one can construct an equivalent normal QRD--0 equation \eqref{GrindEQ__19_} in terms of a basis of solutions of the massless equation associated with \eqref{GrindEQ__17_}:
\be\gamma ^{\mu } D_{\mu } \Psi = \gamma ^{\mu } \left( \partial _{\mu } +\Gamma _{\mu }  \right) \Psi = 0.  
\label{GrindEQ__20_}\ee
This is useful when solutions to the massless equation \eqref{GrindEQ__20_} are known, for which there are many examples in general relativity, including all diagonal metrics and G\"odel type metrics \cite{29.} $-$ \cite{33.}.  Indeed, consider a local similarity transformation $ S $ which takes the Dirac field $ \Psi  $ to $ \widetilde{\Psi } $ and the coefficient fields $ \left( \gamma ^{\mu } , A , \Gamma _{\mu }  \right) $ to $ \left( \widetilde{\gamma }^{\mu } , \widetilde{A} , \widetilde{\Gamma }_{\mu }  \right) $, according to Eqs. \eqref{GrindEQ__5_} and \eqref{GrindEQ__9_}.  Recall that the contracted spin connection matrix $ \widetilde{\Gamma }\equiv \widetilde{\gamma }^\mu \widetilde{\Gamma }_\mu= 0$ if the transformed equation is a  QRD--0 equation.  Then from Eqs. \eqref{GrindEQ__5_} and \eqref{GrindEQ__9_}, we get:

\noindent 
\be\gamma ^{\mu } \left( \partial _{\mu } +\Gamma _{\mu }  \right) S= 0  \label{GrindEQ__21_},\ee

\noindent whereby any four linearly independent solutions to the massless equation \eqref{GrindEQ__20_} can be used to form the columns of the matrix-valued field $ S $.  In this case, an equivalent normal QRD--0 equation can be explicitly and globally constructed.

\noindent \textbf{}

\noindent We can show that any generalized Dirac equation \eqref{GrindEQ__14_}, with very minor conditions imposed on the spacetime metric $ g_{\mu \nu }  $, is equivalent to a normal QRD--0 equation \eqref{GrindEQ__19_}, in an open neighborhood of each spacetime point, by applying local similarity transformations of the first and second kind to the Dirac field $ \Psi  $.  Previously, this so called ``local representation'' of the DFW equation in a curved spacetime has only been discussed in the special case of orthogonal coordinates \cite{3.}, \cite{4.}, \cite{5.}.  Here we generalize it to essentially all Dirac equations:\\

\noindent \textbf{THEOREM 1.}  Let $ {\bf U} $ be any open subset of a spacetime on which local coordinates are defined.  Suppose that the metric components $ g_{\mu \nu }  $ for $ \mu , \nu   =  0, 1, 2, 3 $ in $ {\bf U} $ satisfy $ g_{00}   >  0 $ and the components $ g_{jk}  $ for $ j, k  =  1, 2, 3 $ form a negative definite $ 3\times 3 $ matrix.  Then, for any choice of smooth coefficient fields $ \left( \gamma ^{\mu } , A \right) $ and any choice of covariant derivatives $ D_{\mu } = \partial _{\mu }   +  \Gamma _{\mu }  $ acting on smooth Dirac fields $ \Psi  $ defined on $ {\bf U} $, there exists a smooth local similarity transformation $ S $ of the first kind, composed with a smooth local similarity transformation $ T $ of the second kind, taking  $ \Psi   \to   \left( T\circ S \right)^{-1} \Psi  $, which transforms the generalized Dirac equation with smooth coefficient fields $ \left( \gamma ^{\mu } , A , \Gamma _{\mu }  \right) $ into an equivalent normal QRD--0 equation, in an open neighborhood of each point $ X_{0}   \in   {\bf U} $.\\

\noindent Thus, we may regard the normal QRD--0 equations \eqref{GrindEQ__19_} as canonical forms for all generalized Dirac equations \eqref{GrindEQ__14_}, in open neighborhoods of each spacetime point.  

 The proof of Theorem 1 relies heavily on the theory of linear hyperbolic partial differential equations \cite{7.}, \cite{34.}, \cite{35.}.  Since it is not particularly enlightening beyond the explicit construction given above for the normal case, we will postpone writing out the full proof of Theorem 1 until the Appendix. 

Note that every DFW equation has the normal form \eqref{GrindEQ__17_}.  By Theorem 1, every DFW equation is equivalent to a normal QRD--0 equation (called a ``local representation'' of the DFW equation by Schl\"uter, Wietschorke, and Greiner) which generally is not a DFW equation \cite{3.}, \cite{4.}, \cite{5.}. 
Conversely, we can show that not every normal QRD--0 equation is equivalent to a DFW equation, so that DFW equations are in fact equivalent to only a proper subset of the possible normal QRD--0 equations. \\

\noindent {\bf EXAMPLE.} Consider the flat metric $ g_{\mu \nu }  $ on $ {\bf R}^{4}  $ whose line element $ ds $ in rotating cylindrical coordinates $ \left( t , r , \phi  , z \right) $ has the form:
\be
ds^{2} = (c \,dt)^{2}   -  dr^{2}   -   r^{2} \left( d\phi   +  \omega \, dt \right)^{2}   -   dz^{2} .
\label{GrindEQ__22_}
\ee
Using the orthonormal tetrad \cite{28.} as indicated by the parsing of the metric in Eq. (\ref{GrindEQ__22_}), the DFW equation is given by:
\be
\gamma ^{\mu } D_{\mu } \Psi =  \gamma ^{\mu } \partial _{\mu } \Psi   +  \frac{ 1 }{2r}  \gamma ^{ 1} \Psi  = -i\,\frac{mc}{\hbar }\Psi .
\label{GrindEQ__29_}
\ee

\noindent A simple local similarity transformation of the first kind: 
\be 
S^{-1}   :  \Psi   \mapsto   \Psi '= \sqrt{ r }\,  \Psi   
\label{GrindEQ__30_}
\ee
which is independent of Planck's constant, time independent, and also is independent of the rotation rate $\omega $, transforms the DFW equation \eqref{GrindEQ__29_} into a Dirac equation of the canonical form \eqref{GrindEQ__19_}. 
Note that the canonical Dirac equation \eqref{GrindEQ__19_} does not have the Mashhoon term \cite{9.}, \cite{10.}, whose presence or absence thus depends on the chosen representation of the Dirac field.  See also Ryder \cite{11.}.  

\section{Whitham's Lagrangian Method --- The Main Theorem}\label{WhithamMethod}

\noindent  

\noindent In this section we apply Whitham's Lagrangian method \cite{12.} to derive wave packets for Dirac equations in general curved spacetimes.  Whitham's method preserves the symmetries of the Lagrangian, and in particular, the Whitham wave packet equations conserve the probability current.  We show in this section that generalized de Broglie relations, as well as COW and Sagnac type terms \cite{10.}, emerge from the Whitham equations after transforming each Dirac equation into an equivalent canonical form.  Thus, the normal QRD--0 representations (or canonical forms) of the Dirac equations are the preferred representations to express the generalized de Broglie relations in a curved spacetime.

It is noteworthy that the Whitham approximation places no restriction on Planck's constant $\hbar $.  Indeed, the transformation which takes the Dirac equation to its canonical form, is independent of Planck's constant $\hbar $. This is obvious in the example above, as seen in Eq. (\ref{GrindEQ__30_}).  In fact, this independence is a general fact that follows from Eqs. (\ref{GrindEQ__74_}) and (\ref{GrindEQ__76_}) used in the proof of Theorem 1 in the Appendix.  Thus, the Whitham approximation is not equivalent to the WKB approximation, since the latter does not require any transformation of variables.
\footnote{\
As stated in Section \ref{Canonical Form}, two Dirac equations that are classically equivalent, need not be equivalent with respect to their quantum mechanical energy-momentum operators \cite{25.}, \cite{26.}, \cite{27.}.  Clearly, the Whitham approximation also distinguishes them.  A striking example is a Dirac equation with a Mashhoon term \cite{10.}, \cite{9.}, \cite{11.}.  Applying the Whitham approximation directly to a Dirac equation with a Mashhoon term, without first transforming the Dirac field $\Psi $, does not produce wave packet motion along classical paths.    
}

Including both gravitational and electromagnetic external fields, the generalized Lagrangian \eqref{GrindEQ__8_} for the Dirac equation can be written as follows:
\be\begin{array}{l} {L= L \left( \Psi  ,  \partial _{\mu } \Psi  ,  x^{\mu }  \right)\quad } \\ {} \\ {\quad   =\quad \sqrt{ -g }\,  \frac{ i\hbar c }{2}\,  {\rm [} \Psi ^{+} A\gamma ^{\mu }  \left( D_{\mu } \Psi  \right)  -  \left( D_{\mu } \Psi  \right)^{+} A\gamma ^{\mu }  \Psi   +  \frac{ 2 mc }{\hbar } \, i \Psi ^{+} A\Psi  {\rm ]},} \end{array} 
\label{GrindEQ__33_}\ee
with the covariant derivatives $ D_{\mu }  $ defined by:
\be 
D_{\mu } = \partial _{\mu }   +   \Gamma _{\mu }   +   \frac{ie}{\hbar c}  V_{\mu },  
\label{GrindEQ__34_}
\ee
where $  V_{\mu }  $ are electromagnetic gauge potentials and $ e $ is the electric charge. We can display the interaction terms of the Lagrangian \eqref{GrindEQ__33_} more explicitly by expressing the Lagrangian \eqref{GrindEQ__33_} as a sum of a free and an interaction part as follows: 
\be
\begin{array}{l} {L= \sqrt{ -g }\,  \frac{ i\hbar c }{2}\,  {\rm [} \Psi ^{+} A\gamma ^{\mu }  \left( \partial _{\mu } \Psi  \right)  -  \left( \partial _{\mu } \Psi  \right)^{ +} A\gamma ^{\mu }  \Psi    +   \frac{ 2 mc }{\hbar }  i \Psi ^{+} A\Psi  {\rm ]}} \\ {} \\ {\quad \quad \quad \quad \quad \quad \quad \quad \quad +  \sqrt{ -g } \,  {\rm [} \frac{ i\hbar c }{2}  \Psi ^{+} \left( A\Gamma  - \Gamma ^{+} A \right) \Psi   -  \frac{e}{c}  J^{\mu } V_{\mu }  {\rm ]},} \end{array} 
\label{GrindEQ__35_}
\ee
where $ J^{\mu }   \equiv   c \Psi ^{+} A\gamma ^{\mu } \Psi  $ is the probability current, and $ \Gamma   \equiv   \gamma ^{\mu } \Gamma _{\mu }  $ is the contracted spin connection matrix, and also noting that since $ A $ is a Hermitizing matrix for the Dirac matrices $ \gamma ^{\mu } $, we have from Eq. \eqref{GrindEQ__4_}:  

\noindent 
\be\Gamma ^{+} A= \Gamma _{\mu } ^{+} \gamma ^{\mu +} A= \Gamma _{\mu } ^{+} A\gamma ^{\mu }.  
\label{GrindEQ__36_}
\ee
Since we can first transform any Dirac equation into a normal QRD-0 equation (or canonical form) as stated in Theorem 1 of Section \ref{Canonical Form}, we can transform the fields in this Lagrangian so that $ \nabla _{\mu } \left( A\gamma ^{\mu } \right)  =  0 $ and $ \Gamma   \equiv   \gamma ^{\mu } \Gamma _{\mu }   =  0 $.  Thus, instead of the Lagrangian \eqref{GrindEQ__33_}, we may substitute an equivalent canonical Dirac Lagrangian:

\noindent  
\be\begin{array}{l} {L= \sqrt{ -g } \, \frac{ i\hbar c }{2}\,  {\rm [} \Psi ^{+} A\gamma ^{\mu }  \left( \partial _{\mu } \Psi  \right)  -  \left( \partial _{\mu } \Psi  \right)^{ +} A\gamma ^{\mu }  \Psi    +   \frac{ 2 mc }{\hbar }  i \Psi ^{+} A\Psi  {\rm ]}} \\ {} \\ {\quad \quad \quad \quad \quad \quad \quad \quad \quad -  \sqrt{ -g } \,  \frac{e}{c} \, J^{\mu } V_{\mu },} \end{array} 
\label{GrindEQ__37_}
\ee
Clearly, the normal QRD-0 equation \eqref{GrindEQ__19_} is derived from the Euler-Lagrange equations of the Lagrangian \eqref{GrindEQ__37_} by setting the electromagnetic gauge potentials $  V_{\mu }  $ equal to zero and using the normality condition: $ \nabla _{\mu } \left( A\gamma ^{\mu }  \right)  =  0 $.

For Whitham's method, we set $ \Psi = \chi  e^{i\theta }  $ where $ \chi = \chi \left(X\right) $ is also a complex wave function with four components, and $ \theta = \theta \left(X\right) $ is a real phase at each point $ X $ in the spacetime.  Whitham's method assumes that $ \chi  $ is slowly changing compared to the rapidly changing phase $ \theta $, so that we may obtain approximate wave packet solutions to the Dirac equations by neglecting $ \partial _{\mu } \chi    $ with respect to  $\left( \partial _{\mu } \theta  \right) \chi  $  in the Lagrangian.  Substituting the wave function $ \Psi = \chi  e^{i\theta }  $ into the Lagrangian \eqref{GrindEQ__37_} and using this approximation, we get: 

\noindent 
\be L\quad  =\quad c \sqrt{ -g }   \left[ \left( -\hbar {\kern 1pt} \partial _{\mu } \theta   -  \frac{e}{c}  V_{\mu }  \right)  \chi ^{+} A\gamma ^{\mu } \chi    -   mc \chi ^{+} A \chi  \right].  
\label{GrindEQ__38_}
\ee
In Whitham's method, this Lagrangian governs the wave packet motion. Clearly the Lagrangian (\ref{GrindEQ__38_}) is invariant under the global gauge symmetry $\theta \rightarrow \theta + \tau $, where $\tau  $ is a real constant. This leads by Noether's theorem to the conservation of a current. In Section \ref{Current} we will derive explicitly the conservation of the probability current for this Lagrangian. 
Thus, our goal in this section is to derive the Euler-Lagrange equations for the Lagrangian (\ref{GrindEQ__38_}), which by change of field variables leads to the following main theorem of this paper.

\noindent \textbf{}

\noindent \textbf{THEOREM 2.}  Define a four-vector field $ u^{\mu }  $ and a scalar field  $ J $, related to the amplitude $ \chi  $ and phase $ \theta  $ of the wave function $ \Psi = \chi  e^{i\theta }  $ as follows:

\noindent 
\be\begin{array}{l} {u_{\mu } \equiv -\frac{\hbar }{mc}  \partial _{\mu } \theta    -   \frac{e}{ mc^{2} }  V_{\mu }, } \\ {} \\ {u^{\mu } \equiv g^{\mu \nu } u_{\nu },} \\ {} \\ {J\equiv c \chi ^{+} A \chi .} \end{array} 
\label{GrindEQ__39_}    \ee

\noindent 

\noindent i)  Then the Euler-Lagrange equations for the Whitham Lagrangian \eqref{GrindEQ__38_} imply the following equations for the fields $ u^{\mu }  $ and $ J $ in a curved spacetime: 

\noindent 
\be\begin{array}{l} {g_{\mu \nu }  u^{\mu } u^{\nu }  = 1,} \\ {} \\ {\nabla _{\mu } \left( Ju^{\mu }  \right)= 0,} \\ {} \\ {u_{\mu } = g_{\mu \nu } u^{\nu } ,} \\ {} \\ {\nabla _{\mu } u_{\nu }   -  \nabla _{\nu } u_{\mu } = -\frac{e}{ mc^{2}  } F_{\mu \nu }, } \end{array} 
\label{GrindEQ__40_}  \ee  

\noindent where $ F_{\mu \nu } \equiv \nabla _{\mu } V_{\nu }   -  \nabla _{\nu } V_{\mu }  $ is the electromagnetic field tensor.  

\noindent 

\noindent ii)  The four-vector field $ u^{\mu }  $ is a unit velocity field, such that $ J^{\mu }   =  Ju^{\mu }  $ is a conserved probability current.  

\noindent 

\noindent iii)  The integral curves $ x^{\mu } \left(s\right) $ of the four-vector field $ u^{\mu }  $, parameterized by arc-length $ s $, are given by the classical equations:

\noindent  
\bea
& \frac{\displaystyle dx^{\mu }} {\displaystyle ds} = u^{\mu } , \nonumber \\  & \nonumber \\ & \frac{\displaystyle du^{\mu } }{\displaystyle ds}    +   \left\{^{\ \mu} _{\nu \, \rho} \right\} u^{\nu } u^{\rho }  = \frac{\displaystyle e}{\displaystyle  mc^{2} }  F^{\mu } _{\ \ \nu }  u^{\nu }, 
\label{GrindEQ__41_}    \eea

\noindent along which the scalar field $ J $ satisfies:

\noindent 
\be\frac{dJ}{ds} = -J \nabla _{\mu} u^{\mu }. 
\label{GrindEQ__42_}\ee

\noindent 

\noindent The proof and interpretation of Theorem 2 will occupy the rest of Section 3.    

\subsection{Euler-Lagrange Equations for the Wave Packet Lagrangian}

\noindent 

\noindent The Euler-Lagrange equations for the amplitude $ \chi = \chi \left(X\right) $ and phase $ \theta = \theta \left(X\right) $ can be derived from the wave packet Lagrangian $ L $ in Eq. \eqref{GrindEQ__38_} as follows.  First we have from Eq. \eqref{GrindEQ__38_}:

\noindent  
\be\begin{array}{l} {\frac{\partial L}{\partial \chi ^{+} } = \quad c \sqrt{ -g }  \left[ \left( -\hbar \, \partial _{\mu } \theta   -  \frac{e}{c}  V_{\mu }  \right) A\gamma ^{\mu } \chi    -   mc A \chi  \right],} \\ {} \\ {\frac{\partial L}{\partial \left(\partial _{\mu } \theta \right)} = -\hbar c \sqrt{ -g }   \chi ^{+} A\gamma ^{\mu } \chi.} \end{array} 
\label{GrindEQ__43_}\ee

\noindent Then, since no derivatives of $ \chi ^{+}  $ and only derivatives of $ \theta  $ occur in the Lagrangian \eqref{GrindEQ__38_},  we set equal to zero, using Eq. \eqref{GrindEQ__43_}, the following expressions:

\noindent 
\be\begin{array}{l} {0= \frac{\delta  L}{ \delta \chi ^{+} } =\frac{\partial L}{ \partial \chi ^{+} } }  { = c \sqrt{ -g }  \left[ \left(-\hbar  \partial _{\mu } \theta   -  \frac{e}{c}  V_{\mu }  \right) A\gamma ^{\mu } \chi    -   mc A \chi  \right],} \\ {} \\ 0=\frac{\delta  L}{\delta \theta } = \partial _{\mu }  \left(\frac{\partial L}{\partial \left(\partial _{\mu } \theta \right)} \right)= \partial _{\mu }  \left( -\hbar c \sqrt{ -g }   \chi ^{+} A\gamma ^{\mu } \chi  \right), \end{array} \ee

\noindent which then gives the following Euler-Lagrange equations:

\noindent 
\be\begin{array}{l} {\left( -\hbar \, \partial _{\mu } \theta   -  \frac{e}{c}  V_{\mu } \right) A\gamma ^{\mu } \chi } {=  mc A\chi}, \\ {} \\
\partial _{\mu }  \left( c \sqrt{ -g } \, \chi ^{+} A\gamma ^{\mu } \chi  \right) = 0.\end{array} \label{GrindEQ__45_}\ee

\noindent For the wave function $ \Psi = \chi  e^{i\theta }  $, with phase $ \theta  $, the wave covector is $ K_{\mu }   \equiv   \partial _{\mu } \theta  $.  Thus $ \omega   \equiv   -K_{0}   =  -\partial _{0} \theta  $ is the angular frequency of the wave.  We define a four-vector field $ p^{\mu }  $ as follows:

\noindent 
\be p_{\mu } \equiv  -\hbar\,  \partial _{\mu } \theta = -\hbar K_{\mu }.  \label{GrindEQ__46_}
\ee
\noindent It will be shown in Section \ref{Classical-Quantum} that  $ p_{\mu } =-P_{\mu }  $, where $ P_{\mu }  $ are canonical momentum variables.  

Eq. \eqref{GrindEQ__46_} expresses the generalized de Broglie relations $ P_{\mu }   =  \hbar K_{\mu }   $ between the canonical momentum variables $ P_{\mu }  $ and the wave covector $ K_{\mu }  $.  We also define a four-vector velocity field $ u^{\mu }  $ from the usual classical equation with kinetic and potential terms as follows:

\noindent 
\be p^{\mu } = mc u^{\mu }   +  \frac{e}{c}  V^{\mu }.  
\label{GrindEQ__47_}\ee

\noindent From Eq. \eqref{GrindEQ__46_} we have $ \partial _{\mu } p_{\nu }   =  \partial _{\nu } p_{\mu }  $.  Substituting $ \Psi   =  \chi  e^{i\theta }  $ into Eq. \eqref{GrindEQ__2_} we have $ J^{\mu }   =  c \chi ^{+} A\gamma ^{\mu } \chi  $.  We  denote $ \gamma \left( u \right)  \equiv   u_{\mu } \gamma ^{\mu }  $.  Then, Eqs. \eqref{GrindEQ__45_} become:
\bea
\gamma \left( u \right) \chi & = & \chi \label{gamma chi = chi},            
\\ \label{D_mu J^mu = 0} 
\partial _{\mu }  \left( \sqrt{ -g }  J^{\mu }  \right) & = & 0,
\\ \label{rot p = 0} 
\partial _{\mu } p_{\nu }   -  \partial _{\nu } p_{\mu } & = & 0.
\eea 

\noindent 

\noindent The first equation \eqref{gamma chi = chi} is an algebraic eigenvalue equation.  The second equation \eqref{D_mu J^mu = 0} can be written as the covariant conservation of the probability current, $ \nabla _{\mu } J^{\mu } = 0 $, where $ \nabla _{\mu }  $ denotes the Levi-Civita covariant derivatives.  Since $ p_{\mu }   =  -\hbar {\kern 1pt} \partial _{\mu } \theta  $, the third equation \eqref{rot p = 0} expresses the equality of mixed partial derivatives of $ \theta  $.  Furthermore, since the left-hand side of the third equation \eqref{rot p = 0} is an antisymmetric tensor, the partial derivatives $ \partial _{\mu }  $ can be replaced by Levi-Civita covariant derivatives  $ \nabla _{\mu }  $.  Thus, Eqs. \eqref{gamma chi = chi}--\eqref{rot p = 0} become the following covariant equations: 
\bea\label{gamma chi = chi-bis}
\gamma \left( u \right) \chi & = & \chi,  \\
\label{nabla_mu J^mu = 0} 
\nabla _{\mu } J^{\mu } & = & 0, \\
\label{rot p = 0-bis} 
\nabla _{\mu } p_{\nu }   -  \nabla _{\nu } p_{\mu } & = & 0.
\eea
We will show below that these equations, taken together, reduce to a set of quasi-linear partial differential equations describing a scalar density $ J  =  c \chi ^{+} A\chi  $ and the four-vector velocity field $ u^{\mu }  $, whose integral curves are classical relativistic trajectories.  We will further show that certain of these equations give rise to initial conditions, and of the rest, only four equations are independent.  First, let us derive a dispersion relation from the algebraic equation \eqref{gamma chi = chi-bis}.

\subsection{Dispersion Relation}

From the algebraic equation \eqref{gamma chi = chi-bis}, and the anticommutation relation of Dirac gamma matrices \eqref{GrindEQ__7_}, we have:
\be\left( g^{\mu \nu } u_{\mu } u_{\nu }  \right) \chi = \gamma \left( u \right)^{ 2}  \chi = \gamma \left( u \right) \chi = \chi . 
\label{GrindEQ__50_}\ee
Equation \eqref{GrindEQ__50_} implies that at any spacetime point where the wave function $ \chi  $ is not zero, the four-vector velocity field $ u^{\mu }  $ satisfies $ u^{\mu } u_{\mu }   =  1 $.  From Eq. \eqref{GrindEQ__47_}, this gives the dispersion relation:
\be g^{\mu \nu }  \left( p_{\mu }   -  \frac{e}{c}  V_{\mu }  \right) \left( p_{\nu }   -  \frac{e}{c}  V_{\nu }  \right) - m^{2} c^{2} =0.
\label{GrindEQ__51_}\ee
Since $ p_{0}   =  \varepsilon /c $, where $ \varepsilon  $ is the energy and $ P_{j}   =  -p_{j}  $ for $ j  =  1 , 2 , 3 $ are momentum variables, Eq. \eqref{GrindEQ__51_} is a quadratic equation for the energy $ \varepsilon  $. 

Let us consider the dispersion relation \eqref{GrindEQ__51_} in the absence of the electromagnetic potentials $ V_{\mu }  $.  We have:
\be 
g^{\mu \nu }  p_{\mu } p_{\nu } = g^{00}  \left( \frac{\varepsilon }{c}  \right)^{2}   +  2g^{0j} p_{j}  \left( \frac{\varepsilon }{c}  \right)  +  g^{jk} p_{j} p_{k} = m^{2} c^{2}.  \label{GrindEQ__52_}
\ee
That is,
\be\varepsilon = \frac{c g^{0j} p_{j}   \pm   c \sqrt{ \left( g^{0j} p_{j}  \right)^{ 2}   -  g^{00}  \left( g^{jk} p_{j} p_{k}   -m^{2} c^{2}  \right) } }{g^{00} }.   
\label{GrindEQ__53_}
\ee

\noindent Choosing positive energy $ \varepsilon   >  0 $ and setting:
\be g^{00} = 1  -  \frac{2\phi }{c^{2} } ,\qquad  g^{0j} = \frac{\phi ^{j} }{c} ,\qquad g^{jk} = -\delta ^{jk}   -  \frac{2\phi ^{jk} }{c^{2} },  
\label{GrindEQ__54_}\ee

\noindent where $ \phi  $, $ \phi ^{j}  $, and $ \phi ^{jk}  $ are gravitational potentials, and $ \delta ^{jk}  $ is the Kronecker delta, equal to one if $ j  =  k $ and equal to zero otherwise, then we have for a non-relativistic approximation, i.e., taking the limit of $ \varepsilon   -  mc^{2}  $ as the speed of light $ c $ goes to infinity in Eq. \eqref{GrindEQ__53_}:
\be\begin{array}{l} {\varepsilon   -  mc^{2} = \frac{cg^{0j} p_{j}   +  \sqrt{ g^{00} m^{2} c^{4}   +  \left( cg^{0j} p_{j} \right)^{ 2}   -  c^{2} g^{00} g^{jk} p_{j} p_{k}  } }{g^{00} }    -   mc^{2} } \\ {} \\ {\quad \quad \quad \quad    =\quad \frac{cg^{0j} p_{j}   +  \sqrt{g^{00} } mc^{2} \sqrt{ 1  +  \frac{\left( cg^{0j} p_{j} \right)^{ 2} }{g^{00} m^{2} c^{4} }   -  \frac{g^{jk} p_{j} p_{k} }{m^{2} c^{2} }  } }{g^{00} }    -   mc^{2} } \\ {} \\ {\quad \quad \quad \quad \approx \quad \frac{cg^{0j} p_{j}   +  \sqrt{g^{00} } mc^{2} \left( 1  +  \frac{\left( cg^{0j} p_{j} \right)^{ 2} }{2g^{00} m^{2} c^{4} }   -  \frac{g^{jk} p_{j} p_{k} }{2m^{2} c^{2} }  \right)}{g^{00} }    -   mc^{2} } \\ {} \\ {\quad \quad \quad = c\frac{g^{0j} }{g^{00} } p_{j}   +  \frac{1}{\sqrt{g^{00} } } \left( mc^{2}   +  \frac{\left( cg^{0j} p_{j} \right)^{ 2} }{2g^{00} mc^{2} }   -  \frac{g^{jk} p_{j} p_{k} }{2m}  \right)   -   mc^{2} } \\ {} \\ {\quad \quad \quad \quad \approx \quad \frac{1}{2m}  \delta ^{jk} p_{j} p_{k}   +   m\phi   +  \phi ^{j} p_{j} }. \end{array} 
\label{GrindEQ__55_}\ee

\noindent That is, the non-relativistic approximation gives the energy as follows; 

\noindent 
\be\varepsilon   -  mc^{2} \quad \approx \quad \frac{1}{2m}  \delta ^{jk} p_{j} p_{k}   +   m\phi   +  \phi ^{j} p_{j} . 
\label{GrindEQ__56_}\ee

\noindent It is straightforward to identify the three energy terms on the right hand side of Eq. \eqref{GrindEQ__56_} as the kinetic energy, a COW potential energy, and a Sagnac potential energy, respectively. 

\subsection{Probability Current and Classical Trajectories}\label{Current}

\noindent 

\noindent Recall that for Whitham's method, we set the wave function $ \Psi = \chi  e^{i\theta }  $ where $ \chi = \chi \left(X\right) $ is also a wave function and $ \theta = \theta \left(X\right) $ is a real phase at each point $ X $ in the spacetime.  Then the  probability current $ J^{\mu }  $ and scalar field $J$ are given by:

\noindent   
\bea 
J^{\mu } \equiv c \Psi ^{+} A\gamma ^{\mu } \Psi = c \chi ^{+} A\gamma ^{\mu } \chi,  \nonumber \\
J \equiv c \Psi ^{+} A\Psi = c \chi ^{+} A\chi.
\label{GrindEQ__57_}
\eea

\noindent From Eq. (\ref{gamma chi = chi-bis}), $\chi $ is a solution of the equation $\gamma \left( u \right) \chi = \chi$, where $u^\mu $ is a unit four-vector field satisfying $u^\mu u_\mu =1$ and where $\gamma (u)\equiv u_\mu\gamma ^\mu  $. From the anticommutation relation of the Dirac gamma matrices in Eq. (\ref{GrindEQ__3_}), we have:
\be\label{u vs gamma(u)}
u^\mu =g^{\mu \nu }u_\nu =\frac{1}{2}\left(\gamma ^\mu\gamma ^\nu+\gamma ^\nu\gamma ^\mu \right )u_\nu =\frac{1}{2}\left [\gamma ^\mu\gamma (u)+\gamma (u)\gamma ^\mu \right ].
\ee
Moreover, using again the definition $\gamma (u)\equiv u_\mu\gamma ^\mu  $, it follows easily from the properties of the hermitizing matrix $A$ [Eq. (\ref{GrindEQ__4_})] that $\chi ^+ A \gamma (u)=\left[ \gamma (u)\chi \right]^+A$. We get thus from Eqs. (\ref{gamma chi = chi-bis}), (\ref{GrindEQ__57_}) and (\ref{u vs gamma(u)}): 
\bea
J u^\mu & = & \frac{c}{2}\,\chi ^+A \left[\gamma ^\mu\gamma (u)+\gamma (u)\gamma ^\mu \right ]\chi  \nonumber \\
& = & \frac{c}{2}\,\chi ^+A\gamma ^\mu \left[\gamma (u)\chi  \right ] + \frac{c}{2}\, \left[\gamma (u)\chi  \right ]^+ A\gamma ^\mu\chi \nonumber \\
& = & \frac{c}{2}\,\chi ^+ A\gamma ^\mu\chi + \frac{c}{2}\,\chi ^+A\gamma ^\mu\chi \nonumber \\
& = & J^\mu .
\eea
That is, $J^\mu =J u^\mu $. Using Eqs. \eqref{GrindEQ__47_} and \eqref{GrindEQ__51_} together with this result, Eqs. \eqref{gamma chi = chi-bis}--\eqref{rot p = 0-bis}  can be written as:      
\bea\label{u normed}
g_{\mu \nu }  u^{\mu } u^{\nu }  & = & 1,\\
\label{u bemol}
u_{\mu } & = & g_{\mu \nu }  u^{\nu }, \\
\label{nabla Ju=0}
 \nabla _{\mu } \left( Ju^{\mu }  \right) & = & 0,\\
\label{rot u = C F}
\nabla _{\mu } u_{\nu }   -  \nabla _{\nu } u_{\mu } & = & -\frac{e}{ mc^{2}  } F_{\mu \nu }, 
\eea
where $ F_{\mu \nu } \equiv \nabla _{\mu } V_{\nu }   -  \nabla _{\nu } V_{\mu }  $ is the electromagnetic field tensor.  Multiply by $ u^{\nu }  $ and contract the index $ \nu  $ on both sides of Eq. \eqref{rot u = C F}.  Then, using Eq. \eqref{u normed} to set $ u^{\nu } \left( \nabla _{\mu } u_{\nu }  \right)  =  0 $, and finally raising the index $ \mu  $, we get:

\noindent 
\be\left( u^{\nu } \nabla _{\nu } \right) u^{\mu }  = \frac{e}{ mc^{2} }  F^{\mu } _{\ \ \nu }  u^{\nu }.  
\label{GrindEQ__59_}\ee

\noindent Consider the integral curves $ x^{\mu } \left(s\right) $ of the four-vector velocity field $ u^{\mu } \left(X\right) $.  Since the four-vector velocity field consists of unit vectors by Eq. \eqref{u normed}, the integral curves $ x^{\mu } \left(s\right) $ are parameterized by arc-length $ s $.  That is, from Eqs. \eqref{u normed} and \eqref{GrindEQ__59_}:    

\noindent 
\bea
& \frac{\displaystyle dx^{\mu }} {\displaystyle ds} = u^{\mu } , \nonumber \\  & \nonumber \\ & \frac{\displaystyle du^{\mu } }{\displaystyle ds}    +   \left\{^{\ \mu} _{\nu \, \rho} \right\} u^{\nu } u^{\rho }  = \frac{\displaystyle e}{\displaystyle  mc^{2} }  F^{\mu } _{\ \ \nu }  u^{\nu }. 
\label{GrindEQ__60_}    \eea

\noindent Note that Eqs. \eqref{GrindEQ__60_} are precisely the classical relativistic equations of a particle of mass $ m $ and electric charge $ e $ in a gravitational field $ g_{\mu \nu }  $ in the presence of an electromagnetic field $ F_{\mu \nu }  $.  Thus, the integral curves of the four-vector velocity field $ u^{\mu } \left(X\right) $, describing the motion of wave packets, coincide with the trajectories of classical relativistic particles.  Note that in the absence of the electromagnetic field $ F_{\mu \nu }  $, the classical trajectories \eqref{GrindEQ__60_} are geodesics of the spacetime.  

Finally, we note from Eq. (\ref{nabla Ju=0}) that along the integral curves of  $ u^{\mu } \left(X\right) $ we have: 
\be\frac{dJ}{ds} = -J \nabla _{\mu } u^{\mu }.  
\label{GrindEQ__61_}
\ee
This completes the proof of Theorem 2.  

Note that the wave packet equations \eqref{GrindEQ__40_} describe a certain congruence of classical trajectories \eqref{GrindEQ__41_} together with a scalar density $ J $ that on each classical trajectory in the congruence satisfies Eq. \eqref{GrindEQ__42_}.  This congruence satisfies certain initial conditions on a spatial submanifold $ {\bf M} $ discussed in the next subsection.

\subsection{Mathematical Structure of the Wave Packet Equations}

Eq. \eqref{u normed}--\eqref{rot u = C F} can be written as follows:  

\bea\label{u normed-bis}
g_{\mu \nu }  u^{\mu } u^{\nu }  & = & 1,\\
\label{u bemol-bis}
u_{\mu } & = & g_{\mu \nu }  u^{\nu }, \\
\label{nabla Ju=0-2}
\frac{\partial  }{ \partial x^{\mu } }  \left( \sqrt{ -g }  Ju^{\mu }  \right) & = & 0,\\
\label{rot u = C F-0j}
\frac{\partial u_{0} }{\partial x^{j} }   -  \frac{\partial u_{j} }{\partial x^{0} } & = & \frac{e}{ mc^{2} }  F_{0j} \qquad  (j =  1 , 2 , 3), \\
\label{rot u = C F-jk}
\frac{\partial u_{j} }{\partial x^{k} }   -  \frac{\partial u_{k} }{\partial x^{j} } & = & \frac{e}{ mc^{2} }  F_{jk}\qquad  (j, k  =  1 , 2 , 3).
\eea
Eqs. \eqref{u normed-bis} and \eqref{u bemol-bis} allow us to solve algebraically for $ u^{0}  $, $ u_{0}  $, and $ u_{j}  $ in terms of  $ u^{j}  $ where $ j  =  1 , 2 , 3 $.  We will show below that Eq. \eqref{rot u = C F-jk} gives merely a set of initial conditions.  Thus, we are left with only four real quasi-linear partial differential equations contained in Eqs.  \eqref{nabla Ju=0-2} and \eqref{rot u = C F-0j}, for the four real fields $ u^{j}  $ and $ J $.  Indeed, from Eq. \eqref{GrindEQ__47_}, we have that \eqref{rot u = C F-0j} and \eqref{rot u = C F-jk} are equivalent to:
\bea\label{rot p  0j = 0}
\frac{\partial p_{j} }{\partial x^{0} } &  =&  \frac{\partial p_{0} }{\partial x^{j} }, \\
\label{rot p  jk = 0}
\frac{\partial p_{j} }{\partial x^{k} } &  = &  \frac{\partial p_{k} }{\partial x^{j} }. 
\eea
It follows from Eq. \eqref{rot p  0j = 0} that Eq. \eqref{rot p  jk = 0} remains true for all time if and only if it is true at an initial time.  That is because from Eq. \eqref{rot p  0j = 0}, we derive:
\be\frac{\partial }{\partial x^{0} } \left(\frac{\partial p_{j} }{\partial x^{k} } \right) =  \frac{\partial ^{2} p_{0} }{\partial x^{j} \partial x^{k} } =  \frac{\partial }{\partial x^{0} } \left(\frac{\partial p_{k} }{\partial x^{j} } \right). 
\label{GrindEQ__65_}\ee

\noindent It follows that Eq. \eqref{rot u = C F-jk} gives a set of initial conditions.  Thus, provided that the initial conditions \eqref{rot u = C F-jk} are satisfied at any initial time --- i.e., on a spatial submanifold $ {\bf M} $--- the wave packet equations \eqref{u normed-bis}--\eqref{rot u = C F-jk} give rise to well-defined solutions.  As previously stated these solutions comprise a congruence of classical trajectories together with a scalar density. 

\subsection{Gordon Decomposition for Dirac Equations in Canonical Form}

In the WKB approximation of the DFW equation in a curved spacetime, classical trajectories are derived from the Gordon decomposition of the probability current $J^\mu = J^\mu _c + J^\mu _s$ into a convection current $J^\mu _c$ and a spin current $J^\mu _s$ \cite{Audretsch1981A}. In this subsection, we will prove the existence of the Gordon decomposition for all normal Dirac equations, noting that both DFW and canonical equations are normal.  We will further show that in the Whitham approximation, the spin current $J^\mu _s$ vanishes, which explains the negligible effect of spin on wave packet solutions.  More specifically, we will show that wave packet solutions of the form $\Psi =\chi e^{i\theta }$, where $\chi $ is slowly changing compared to a rapidly changing phase $\theta $, can only exist if the spin current $J^\mu_s$ is negligible.  As discussed above, this definition of wave packet is independent of the size of Planck's constant $\hbar $.  Indeed, no assumption will be made in this subsection regarding the size of Planck's constant $\hbar $ or the speed of light $c$, both of which we will set equal to one, $\hbar =c=1$.  

For a normal Dirac equation, define the probability current $J^\mu$, the spin current $J^\mu_s$, and the convection current $J^\mu_c$ as follows:      

\bea 
J^{\mu } & \equiv & \Psi ^{+} A\gamma ^{\mu } \Psi,  \nonumber \\
J^{\mu }_s & \equiv & \frac{1}{2m} \left[\Psi ^{+} A\sigma ^{\mu \nu }(D_\nu \Psi) + (D_\nu \Psi)^+A\sigma ^{\mu \nu }\Psi \right ], \nonumber \\
J^{\mu }_c & \equiv & \frac{i}{2m} g^{\mu \nu }\left[\Psi ^{+} A(D_\nu \Psi) - (D_\nu \Psi)^+A\Psi \right ],
\label{Gordon}
\eea
where $\sigma ^{\mu \nu }\equiv \frac{i}{2} \left (\gamma ^\mu \gamma ^\nu -\gamma ^\nu \gamma ^\mu \right )$ are the Dirac spin matrices.  Substituting the normal Dirac equation (\ref{GrindEQ__17_}) written in the following form:
\be\label{Dirac Psi=}
\Psi = \frac{i}{m}\gamma ^\mu D_\mu \Psi 
\ee
into the formula for $J^\mu$ in Eq. (\ref{Gordon}), using the gamma matrix formula $\gamma ^\mu \gamma ^\nu= g^{\mu \nu }-i\sigma ^{\mu \nu }$, and noting the asymmetry $\sigma ^{\mu \nu }=-\sigma ^{\nu \mu }$, we get from Eqs. (\ref{Gordon}) and (\ref{Dirac Psi=}):
\bea
J^{\mu } & \equiv & \Psi ^{+} A\gamma ^{\mu } \Psi =  \frac{i}{2m} \left[\Psi ^{+} A\gamma ^\mu \gamma ^\nu(D_\nu \Psi) - (D_\nu \Psi)^+A\gamma ^\nu \gamma ^\mu\Psi \right ]\nonumber \\
 & = & \frac{i}{2m} \left[\Psi ^{+} A\left(g^{\mu \nu }-i\sigma ^{\mu \nu }\right)(D_\nu \Psi) - (D_\nu \Psi)^+A\left(g^{\mu \nu }+i\sigma ^{\mu \nu }\right)\Psi \right ] \nonumber \\
 & = & J^\mu_c + J^\mu_s.
\label{Gordon2}
\eea
That is, $J^\mu = J^\mu_c + J^\mu_s$, which proves the Gordon decomposition for normal Dirac equations.  

Note that the probability current $J^\mu = J^\mu_c + J^\mu_s$ is covariantly conserved. Consequently, the probability density current $\sqrt{-g} J^\mu$ is conserved.  In general, the currents $J^\mu_c$ and $J^\mu_s$ are not separately covariantly conserved, unless the coefficient fields $(\gamma ^\mu ,A)$ are covariantly constant. However, it is worthy to note that for Dirac equations transformed into canonical form, we may replace the covariant derivatives $D_\mu$, including electromagnetic field potentials $V_\mu $ as in Eq. (\ref{GrindEQ__34_}), with the Levi-Civita covariant derivatives $\nabla _\mu +ieV_\mu $.
\footnote{\
Note that the  probability current $J^\mu =\Psi ^+A\gamma ^\mu \Psi $ is invariant under local similarity transformations $S$ defined in Eq. (\ref{GrindEQ__5_}), so that when transforming a Dirac equation into canonical form, it is only $J^\mu _s$ and $J^\mu _c$ which change their form.   
}
Then the spin current $J^\mu_s$ and the convection current $J^\mu_c$ become:      
\bea 
J^{\mu }_s & = & \frac{1}{2m} \left[\Psi ^{+} A\sigma ^{\mu \nu }(\partial _\nu \Psi) + (\partial _\nu \Psi)^+A\sigma ^{\mu \nu }\Psi \right ], \nonumber \\
J^{\mu }_c & = & \frac{i}{2m} g^{\mu \nu }\left[\Psi ^{+} A(\partial  _\nu +ieV_\nu)  \Psi - \left((\partial  _\nu +ieV_\nu ) \Psi\right )^+A\Psi \right ].
\label{Gordon3}
\eea
Note that the convection current $J^{\mu }_c$ in Eq. (\ref{Gordon3}) closely resembles the spin zero current of the Klein-Gordon equation in the presence of both gravitational and electromagnetic external fields \cite{28.}, whereas the spin motion resides in the spin current $J^{\mu }_s$ \cite{Audretsch1981A}. Recall that we are using Whitham's approximation, for which we set $\Psi =\chi e^{i\theta }$ and neglect $\partial _\mu \chi $ with respect to $(\partial _\mu \theta ) \chi $ [see before Eq. (\ref{GrindEQ__38_})]. Using the definitions in Eq. (\ref{GrindEQ__39_}), this gives us
\be\label{Whitham to D Psi}
(\partial  _\nu +ieV_\nu)  \Psi \approx i (\partial  _\nu \theta +eV_\nu)  \Psi=-imu_\nu \Psi .
\ee
Then we use the definitions in Eq. (\ref{GrindEQ__39_}) together with Eqs. (\ref{Gordon3}) and (\ref{Whitham to D Psi}) to obtain the following formulas for the spin current and the convection current: 
\bea 
J^{\mu }_s & \approx  & -\frac{i}{2} \left[\Psi ^{+} A\sigma ^{\mu \nu }u_\nu \Psi - \Psi^+A\sigma ^{\mu \nu }u_\nu \Psi \right ]=0, \nonumber \\
J^{\mu }_c & \approx  & \frac{1}{2} g^{\mu \nu }u_\nu \left[\Psi ^{+} A\Psi + \Psi ^{+} A\Psi \right ]=Ju^\mu .
\label{Gordon4}
\eea
Thus, in the Whitham approximation, the spin current $J^{\mu }_s$ vanishes and the convection current $J^\mu_c = Ju^\mu$. Therefore, the probability current $J^\mu$ equals the convection current $J^\mu_c$.

\section{The Classical-Quantum Correspondence}\label{Classical-Quantum}

We proved in Theorem 2 that the solutions of the Whitham approximation to the Dirac equation, Eq. \eqref{GrindEQ__40_}, consist of a four-velocity vector field $ u^{\mu }$ whose integral curves are classical trajectories, and a scalar field $ J $ representing a conserved particle density $-$ see Eqs. \eqref{GrindEQ__41_} and \eqref{GrindEQ__42_}.  From Whitham's approximation of the Dirac equation, we have derived the dispersion relation \eqref{GrindEQ__51_}, the motion of wave packets along classical trajectories \eqref{GrindEQ__41_}, conservation of the probability current \eqref{GrindEQ__42_}, as well as the generalized de Broglie relations $ P_{\mu }   =  \hbar K_{\mu }   $ in Eqs. \eqref{GrindEQ__46_} and \eqref{GrindEQ__47_}.   

We will conclude this paper by summarizing results from a previous analysis of the ``classical-quantum correspondence'' based on the dispersion relation \eqref{GrindEQ__51_} alone, which can be applied to the Dirac equation \cite{15.}.  We will interpret the fact that the integral curves of the four-velocity field $ u^{\mu }  $ are classical trajectories as a ``geometrical optics limit'' of the Dirac equation.  However, as we have seen, unlike the WKB limit, this ``geometrical optics limit'' is one that places no restriction on Planck's constant.  Whitham's Lagrangian method, applied to the Dirac Lagrangian (\ref{GrindEQ__33_}) transformed into equivalent canonical form (\ref{GrindEQ__37_}), has also many advantages over using the dispersion relation \eqref{GrindEQ__51_} alone as the starting point for a wave packet approximation \cite{12.}.  Whitham's method preserves the conservation laws inherent in the starting Lagrangian \eqref{GrindEQ__37_}, and in particular, the Whitham wave packet equations conserve the probability current $ J^{\mu }   =  Ju^{\mu }  $.  

To any linear partial differential equation for scalar wave functions $ \Psi  $, of the form: 

\noindent 
\be a\left( X \right) \Psi   +  \sum _{n = 1}^{d}  a^{^{\mu _{_{1} }  \cdots \cdots  \mu _{n} } } \left( X \right) \frac{\partial ^{n} \Psi }{\partial x^{\mu _{ 1} } ...... \partial x^{\mu _{ n} } } = 0\label{GrindEQ__66_}\ee

\noindent (summing over coordinate indices $ \mu _{r}   =  0, 1, 2, 3 $ for $ r  =  1, 2, 3,  \cdots   , n $ and over the index $ n  =  1, 2, 3,  \cdots   , d $) where the coefficient fields $ a\left( X \right) $ and $ a^{^{\mu _{_{1} }  \cdots \cdots  \mu _{n} } } \left( X \right) $ depend on the spacetime point $ X $$-$  one may associate its dispersion polynomial $ \Pi _{X} \left( K \right)$.  That is to say, a polynomial function of covector fields $ K_{\mu }  $ at each fixed spacetime point $ X $ is given by:  
\be
\Pi _{X} \left( K \right)= a\left( X \right)   +  \sum _{n = 1}^{d}  i^{n}  a^{^{\mu _{{1} }  \cdots \cdots  \mu _{ n} } } \left( X \right)  K_{\mu _1 }    ......  K_{\mu _n}. 
\label{GrindEQ__67_}
\ee

The dispersion relation is thereby obtained from the polynomial equation $ \Pi _{X} \left( K \right)= 0 $ at each fixed spacetime point $ X $ by solving for the time component $ K_{0}  $.  Applications of this one-to-one correspondence are discussed in Ref. \cite{15.}.

This applies also if the wave functions $ \Psi  $ have $ m $ components and the coefficient fields $ a\left( X \right) $ and $ a^{^{\mu _{_{1} }  \cdots \cdots  \mu _{n} } } \left( X \right) $  are $ m\times m $ matrices, as is the case for the Dirac equation \cite{21.}, \cite{36.}.  Note that in the matrix case, the dispersion relation is obtained from the scalar polynomial equation $ \det  \Pi _{X} \left( K \right)= 0 $.  Consider a dispersion polynomial \eqref{GrindEQ__67_} where $ a\left( X \right) $ and $ a^{^{\mu _{_{1} }  \cdots \cdots  \mu _{n} } } \left( X \right) $  are $ m\times m $ matrices. By solving $ \det  \Pi _{X} \left( K \right)= 0 $ for the component $ K_{0}  $ we get a dispersion relation:

\noindent        
\be \omega = W\left( {\bf k} , {\bf x} {\bf ,} t \right)  
\label{GrindEQ__68_}\ee

\noindent expressing the angular frequency $ \omega   \equiv   -c\,K_{0}  $ as a function of the spatial wave covector $ {\bf k}  =  \left( {\it K}_{{\rm 1}}  , K_{{\rm 2}}  , K_{{\rm 3}}  \right) $, the spatial coordinates $ {\bf x}  =  \left( x^{1}  , x^{2}  , x^{3}  \right) $, and time $ t $, together with the auxiliary equations \cite{12.}, \cite{15.}: 

\noindent 
\bea \frac{\displaystyle \partial K_{j} }{\displaystyle \partial t }   +  \frac{\displaystyle \partial \omega  }{\displaystyle \partial x^{j} } = 0, \nonumber \\ \nonumber \\  \frac{\displaystyle \partial K_{j} }{\displaystyle \partial x^{k} }   -  \frac{\displaystyle \partial K_{k} }{\displaystyle \partial x^{j} } = 0,  
\label{GrindEQ__69_}\eea

\noindent for $ j, k  =  1 , 2 , 3 $.  In general, none or multiple such dispersion relations \eqref{GrindEQ__68_} can be derived as distinct real roots of the polynomial equation $ \det  \Pi _{X} \left( \omega  , {\bf k} \right)= 0 $ when solving for $ \omega  $.  Assuming at least one real root, let us choose one of them to be $ \omega = W\left( {\bf k} , {\bf x} {\bf ,} {\rm t} \right) $.  Then from Eqs. \eqref{GrindEQ__68_} and \eqref{GrindEQ__69_} one derives the Hamiltonian system as in Ref. \cite{15.}, Sect. 2.2: 
\be
\frac{d K_j}{d t}= -\frac{\partial  W}{\partial x^j},\qquad \frac{d x^j}{d t}= \frac{\partial  W}{\partial K_j}.
\label{GrindEQ__70_}\ee

\noindent Noting in Eq. \eqref{GrindEQ__68_} that $ W  \equiv   -c\,K_{0}  $, and recalling the generalized de Broglie relations $ P_{\mu }   \equiv   -p_{\mu }   =  \hbar  K_{\mu }  $ derived from wave packet motion in Eqs. \eqref{GrindEQ__46_} and \eqref{GrindEQ__47_}, we see that Eq. \eqref{GrindEQ__70_} leads us to define the Hamiltonian $ H  \equiv   \hbar  W $ and the momentum variables $ P_{j}   \equiv   \hbar  K_{j}  $ for $ j  =  1 , 2 , 3 $, whereby a system of classical point particle trajectories emerges as follows:  Indeed, solving the dispersion equation \eqref{GrindEQ__51_} for the energy $ \varepsilon = H\left( {\bf p} , {\bf x} {\bf ,} {\rm t} \right) $ where $ {\bf p}  =  \left( P_{{\rm 1}}  , P_{{\rm 2}}  , P_{{\rm 3}}  \right) $ is equivalent to solving it for the angular frequency $ \omega = W\left( {\bf k} , {\bf x} , t \right) $ as in Eq. \eqref{GrindEQ__68_}.  Then Eq. \eqref{GrindEQ__70_} is equivalent to:

\noindent 
\be\frac{d P_j}{d t}= -\frac{\partial  H}{\partial x^j},\qquad \frac{d x^j}{d t}= \frac{\partial  H}{\partial P_j}. 
\label{GrindEQ__71_}\ee

\noindent One can show that the trajectories associated with the Hamiltonian $ H $, that is, the solution trajectories of the Hamiltonian equations \eqref{GrindEQ__71_}, are identical to the solution trajectories of the Euler-Lagrange equations deduced from the well known Lagrangian for classical point particles in a background of electromagnetic and gravitational  fields, which is given by \cite{37.}, \cite{38.}:

\noindent 
\be\ell = -mc \sqrt{ g_{\mu \nu } \frac{dx^{\mu } }{d\xi  } \frac{dx^{\nu } }{d\xi  }  }    -   \frac{e}{c}  V_{\mu } \frac{dx^{\mu } }{d\xi  },   \label{GrindEQ__72_}\ee

\noindent where $ \xi  $ is an arbitrary parameter for the classical trajectory $ x^{\mu } \left( \xi  \right) $.  To show this one first applies an inverse Legendre transformation
\footnote{   The Legendre transformation is its own inverse \cite{38.}, pages 563-565.  Thus, an inverse Legendre transformation is also a Legendre transformation.  
}
 to the Hamiltonian $ H  =  H\left( {\bf p} , {\bf x} {\bf ,} {\rm t} \right) $ to obtain a traditional Lagrangian $ L \left( {\bf x} ,  \frac{d{\bf x}}{dt}  ,  t \right) $ and then, as in Ref. \cite{38.}, pages 267-271, one generalizes the trajectory parameter to be an arbitrary parameter $ \xi  $, instead of the coordinate time $ t $.  It is straightforward to check that the equations for the classical trajectories \eqref{GrindEQ__41_} are the Euler$-$Lagrange equations for the Lagrangian \eqref{GrindEQ__72_}.  Thus, the dispersion relation \eqref{GrindEQ__68_} and the auxiliary equations \eqref{GrindEQ__69_} give rise to the classical point particle trajectory equations \eqref{GrindEQ__41_}, as is also the case for the integral curves of the Whitham equations \eqref{GrindEQ__40_}.  

Similar to the previous analysis of the classical-quantum correspondence (\cite{15.}, Sect. 2.3), it is in the ``geometrical optics limit'' that the solutions of each Dirac equation transformed into equivalent canonical form satisfy the dispersion equation \eqref{GrindEQ__51_}.  Indeed, Whitham's approximation:  $ \partial _{\mu } \chi    \ll \left( \partial _{\mu } \theta  \right) \chi  $, which we applied in Section \ref{WhithamMethod} to the Dirac Lagrangian \eqref{GrindEQ__37_}, is one way of defining precisely this limit.  However, the classical Hamiltonian equations \eqref{GrindEQ__71_}, which are based solely on the dispersion relation, give only part of the Whitham equations \eqref{GrindEQ__40_}.  In addition to providing equations \eqref{GrindEQ__41_} equivalent to the Hamiltonian equations \eqref{GrindEQ__71_}, the Whitham equations \eqref{GrindEQ__40_} preserve the symmetries of the Dirac Lagrangian \eqref{GrindEQ__37_}, and provide for the conservation of the probability current $ J^{\mu }   =  Ju^{\mu }  $, which is a property inherited from the exact Dirac equations in a curved spacetime.  

\appendix

\section{Appendix: Proof of Theorem 1}

To prove Theorem 1 of Section \ref{Canonical Form} we will use the following corollary of a deep theorem of linear hyperbolic partial differential equations: \\  

\noindent \textbf{THEOREM 0.}  Let $ M_{1} , M_{2} , \cdots  , M_{n}  $ be complex $ d\times d $ matrix functions that depend smoothly on $ n+1 $ independent real variables $ t , x_{1}  , x_{2}  ,  \cdots  , x_{n}  $ in a slab $ -T  \le   t  \le   T $, $ x  \in   {\bf R}^{{\rm n}}   $, denoted as $ {\bf I}\times {\bf R}^{{\rm n}}  $.  Suppose that $ M_{0} , M_{1} , M_{2} , \cdots  , M_{n}  $ are Hermitian matrices, and furthermore assume that $ M_{0}  $ is positive definite.  Let $ F  =  F\left( S \right) $ be a homogeneous linear function of complex $ d\times d $ matrices $ S $, as well as having explicit dependence on $ t , x_{1}  , x_{2}  ,  \cdots  , x_{n}  $.  Then the complex linear hyperbolic system:
\be 
M_{0} \frac{\partial S}{\partial t}   +  \sum _{j=1}^{n} M_{j} \frac{\partial S}{ \partial x_{j} } = F\left( S \right) 
\label{GrindEQ__73_}
\ee
has a smooth $ d\times d $ complex  matrix valued solution $ S {\bf :}  {\bf I}\times {\bf R}^{{\rm n}}   \to    {\it M}\left( {\it C} {\it ,} {\it d} \right) $ which equals the identity matrix at $ t  =  0 $, as its prescribed smooth initial data.\\

\noindent \textbf{Proof:}  Theorem 0 is Corollary 3 in Ref. \cite{7.}, Sect. 6, which is based on a theorem of Lax \cite{34.}  (see also Ref. \cite{35.}).  \\    

\noindent \textbf{Proof of Theorem 1 of Section 2.}\\

\noindent First note from Eq. \eqref{GrindEQ__4_} that the matrices $ B^{\mu }   \equiv   A\gamma ^{\mu }  $ are Hermitian matrices.  Then, note that the conditions stated in Theorem 1 for the metric components $ g_{\mu \nu }  $ imply that  $ B^{0}   \equiv   A\gamma ^{0}  $ is a positive definite matrix by Theorem 6 of Ref. \cite{22.}, Appendix B.  By Theorem 3 of Ref. \cite{8.}, Sect. 3.4, Eq. (54), a local similarity transformation $ T $ of the second kind, takes a generalized Dirac equation of the form \eqref{GrindEQ__14_} into a Dirac equation of the normal form \eqref{GrindEQ__17_}, if and only if $ T $ satisfies the following partial differential equation:    \textbf{    }

\be B^{\mu } D_{\mu } T= -\frac{1}{2}  \left( D_{\mu } B^{\mu }  \right) T .
\label{GrindEQ__74_}\ee

\noindent If we can solve Eq. \eqref{GrindEQ__74_} for such a local similarity transformation $ T $, then the transformed coefficient fields will be as in Eqs. \eqref{GrindEQ__5_} and \eqref{GrindEQ__11_}:

\noindent 
\be\begin{array}{l} {\widetilde{\gamma }^{\mu } = T^{-1} \gamma ^{\mu }  T,} \\ {} \\ {\widetilde{A}\quad  =\quad T^{+} A T,} \\ {} \\ {\widetilde{\Gamma }_{\mu } = \Gamma _{\mu } .} \end{array} 
\label{GrindEQ__75_}\ee

\noindent Now from Eq. \eqref{GrindEQ__21_}, a local similarity transformation $ S $ of the first kind, takes a Dirac equation of the normal form \eqref{GrindEQ__17_} into a normal QRD--0 equation \eqref{GrindEQ__19_}, if and only if $ S $ satisfies the following partial differential equation: 

\noindent 
\be\widetilde{B}^{\mu } \partial _{\mu } S= -\widetilde{B}^{\mu } \widetilde{\Gamma }_{\mu }  S, 
   \label{GrindEQ__76_}\ee

\noindent where $ \widetilde{B}^{\mu }   \equiv   \widetilde{A}\widetilde{\gamma }^{\mu }  $.  The matrices $ B^{\mu }  $ and $ \widetilde{B}^{\mu }  $ are  Hermitian, and moreover, the matrices $ B^{0}  $ and $ \widetilde{B}^{0}  $ are positive definite.  Thus, the two systems \eqref{GrindEQ__74_} and \eqref{GrindEQ__76_} have same form as the complex linear hyperbolic system in Eq. \eqref{GrindEQ__73_}.  

Let $ \chi   {\bf :}  {\bf U}  \to   {\bf R}  \times   {\bf R}^{{\bf 3}}  $, mapping $ X  \to   \left( t , {\bf x} \right) $, be the coordinate chart that we assume to be defined on $ {\bf U} $.  Then, consider the projection map $ \pi   {\bf :}  {\bf R}  \times   {\bf R}^{{\bf 3}}   \to   {\bf R} $ taking $ \left( t , {\bf x} \right)  \to   t $.  Let $ X_{0}   \in   {\bf U} $ and let $ t_{0}   =  \pi \circ \chi \left( X_{0}  \right)  \in   {\bf R} $.  Let $ {\bf M}  =  \left( \pi \circ \chi  \right)^{-1} \left( t_{0}  \right) $.  Note that $ X_{0}   \in   {\bf M}  \subset   {\bf U} $.  It will suffice to prove that there exist nonsingular solutions $ T $ and $ S $ of the systems \eqref{GrindEQ__74_} and \eqref{GrindEQ__76_} that are both defined in a common open neighborhood $ \widetilde{{\bf W}} $ of $ X_{0} \in   {\bf U}$.

By Theorem 0, the Cauchy problem for \eqref{GrindEQ__74_} with the smooth initial data $ T|_{ {\bf M}}   =  {\bf 1}_{{\bf 4}} $ has a smooth solution $ T $ in an open neighborhood  $ {\bf W}^{{\bf \# }}  $ of $ X_{0}  $.  Denote by $ {\bf W} $ the open subset of $ {\bf W}^{{\bf \# }}  $ in which $ T $ is a nonsingular matrix so that $ T^{-1}  $ exists.  Note that $ X_{0}   \in   {\bf W} $ since $ X_{0}   \in   {\bf M} $ and $ T|_{ {\bf M}}   =  {\bf 1}_{{\bf 4}}  $.  Thus, $ {\bf W} $ is an open neighborhood of $ X_{0}  $ such that the local similarity transformation $ T $ (and its inverse $ T^{-1}  $) is well defined on $ {\bf W} $, and hence, from Eq. \eqref{GrindEQ__75_}, the complex linear hyperbolic system \eqref{GrindEQ__76_} is well defined on $ {\bf W} $.

By Theorem 0, the Cauchy problem for \eqref{GrindEQ__76_} with the smooth initial data $ S|_{ {\bf M}\bigcap {\bf W}}   =  {\bf 1}_{{\bf 4}}  $ has a smooth solution $ S $ in an open neighborhood $ \widetilde{{\bf W}}^{{\bf \# }}   \subset   {\bf W} $ of $ X_{0}  $.  Denote by $ \widetilde{{\bf W}} $ the open subset of  $ \widetilde{{\bf W}}^{{\bf \# }}  $ in which $ S $ is a nonsingular matrix so that $ S^{-1}  $ exists.  Note that $ X_{0}   \in   \widetilde{{\bf W}} $ since $ X_{0}   \in   {\bf M}\bigcap {\bf W} $ and $ S|_{ {\bf M}\bigcap {\bf W}}   =  {\bf 1}_{{\bf 4}}  $.  Thus, $ \widetilde{{\bf W}}  \subset   \widetilde{{\bf W}}^{{\bf \# }}   \subset   {\bf W} $ is an open neighborhood of $ X_{0}  $ such that both the local similarity transformation $ S $ and the local similarity transformation $ T $ (and their inverses $ S^{-1}  $ and $ T^{-1}  $) are well defined on $ \widetilde{{\bf W}} $, and therefore the local similarity transformation $ T\circ S $ (and its inverse $ S^{-1} \circ   T^{-1}  $) is also well defined on $ \widetilde{{\bf W}} $.  \textbf{Q. E. D.}\\

\vspace{2mm}
\end{document}